\documentclass[twocolumn,journal]{IEEEtran}

\usepackage{graphicx}
\usepackage{amsmath}
\usepackage{amssymb}
\usepackage{balance}

\usepackage{amsthm}
\usepackage{cite}
\usepackage{float}
\usepackage{color}
\usepackage{hyperref}
 \newtheorem{theorem}{Theorem}

\begin{document}

\title{Cooperative Cognitive Relaying Under Primary and Secondary Quality of Service Satisfaction}
\author{Ahmed~El~Shafie,~\IEEEmembership{Member,~IEEE,}
        Tamer~Khattab,~\IEEEmembership{Member,~IEEE}
        \thanks{Part of this work has been accepted in the IEEE International Symposium on Personal, Indoor and Mobile Radio Communications (PIMRC), 2014.}
\thanks{A. El Shafie is with Wireless Intelligent Networks Center (WINC), Nile University, Giza, Egypt. e-mail: ahmed.salahdeen@yahoo.com}
\thanks{T. Khattab is with Electrical Engineering, Qatar University, Doha, Qatar. email: tkhatta@ieee.org}
\thanks{This research work is supported by Qatar National Research Fund (QNRF) under grant number NPRP 6-1326-2-532.}}
\date{}
\maketitle
\begin{abstract}
This paper proposes a new cooperative protocol which involves cooperation between primary and secondary users. We consider a cognitive setting with one primary user and multiple secondary users. The time resource is partitioned into
discrete time slots. Each time slot, a secondary user is scheduled for transmission according to time division multiple access, and the remainder of the secondary users, which we refer to as secondary relays, attempt to decode the primary packet. Afterwards, the secondary relays employ cooperative beamforming
to forward the primary packet and to provide protection to the
secondary destination of the secondary source scheduled for transmission from interference. We characterize the diversity-multiplexing tradeoff of the primary source under the proposed protocol. We consider certain quality of service for each user specified by its required throughput. The optimization problem is stated under such condition. It is shown that the optimization problem is linear and can be readily solved. We show that the sum of the secondary required throughputs must be less than or equal to the probability of correct packets reception.
\end{abstract}

\section{Introduction}
 %
 %
\IEEEPARstart{C}\small{ognitive} Radio (CR) is a promising technology to
improve the utilization of spectrum bands.
The main idea was established by enabling
opportunistic spectrum sharing.
The CR users (or secondary
users) dynamically utilize the licensed frequency
spectrum of primary licensed systems under the condition that the interference
to primary users remains below a certain threshold.

Beamforming is an emerging and efficient technology
that enables concurrent transmissions of different nodes in the network. Recently, it has been applied to cognitive radio networks; a network with set of primary users and secondary users \cite{a2,a3,a4}. The importance of beamforming  is due to the fact that it can support multiple user streams on separate spatial paths at the same spectrum
simultaneously \cite{a5,a6}.
A set of distributed nodes can perform beamforming  by utilizing
a `virtual' antenna array that can be created by a set of
nodes in cooperative relaying networks \cite{a7,a8}. Thus,
a distributed beamformer can be created by carefully selecting
the beamforming  weight in each relay node.

Performing beamforming without causing interference at certain node is referred
to as zero-forcing beamforming \cite{a6}. To the best of our knowledge,
however, the problem of designing a distributed zero-forcing
beamformer in a relay-assisted cognitive network to enable one of the secondary users to utilize the spectrum concurrently with the other secondary users which relay a primary packet using distributed beamforming has not
been addressed. It is worth pointing out that the
proposed beamforming, formed by multiple secondary relays, can achieve cooperative diversity gain \cite{b0} for primary users, and at the same time create a beamformer to null the interference to the destination of the active secondary users, i.e., secondary users scheduled for transmission. We emphasize the following, as mentioned in \cite{liu2012delay}, most existing work on applying beamforming in cognitive radio networks did not consider node cooperation
\cite{b1,b2,b3,b4}. On the other hand, many of existing work
on cooperative/distributed beamforming has
rarely considered its application in cognitive radio networks \cite{b5,b6,liu2012delay}.

In \cite{liu2009distributed}, the authors proposed a distributed zero-forcing beamforming
approach to increase the opportunistic spectrum access for the secondary users in
cognitive radios networks. Specifically, the secondary source accesses temporal
spectrum holes to broadcast a message to a set of relays,
which in turn form a distributed zero-forcing beamformer and start a simultaneous transmission with the active primary users,
without causing interference to any of the primary receiving nodes. In \cite{liu2012delay} and \cite{liu2012cooperative}, the same authors of \cite{liu2009distributed} considered a relaying cooperative network,
in which a set of relays equipped with finite-sized buffers were assumed to
aid the secondary source transmission using cooperative beamforming. The authors showed the improvement of the quality of service (QoS) of the secondary source in terms of packets queueing delays.

In this paper, we consider a cognitive network with one primary user and a set of secondary users. Each user has certain throughput requirement. We propose a distributed beamforming method to enable
simultaneous transmissions of secondary users with active primary users, while
ensuring no interference to secondary users in a relay-assisted manner.
Specifically, the primary user broadcasts its packet to its destination and a set of secondary users which temporary operate as relay stations for the primary user. One of the secondary users is assigned to access the time slot simultaneously with the other secondary users which form a distributed zeroforcing
beamformer to capable of forwarding the primary message. The zero-forcing beamformer, designed to maximize
the received signal-to-interference-plus-noise ratio (SINR) at
the primary destination while completely eliminating the interference to the destination of the active
secondary user, is successfully formed each time slot through a method of orthogonal
projection. We analyze the outage
probability of nodes under the assumption of slow fading channels
between links. We consider two schemes based on the state of connectivity of the primary direct link. Through
theoretical analysis, we find that the spatial diversity order of
our proposed scheme is equal to the total number of secondary relays minus one or minus two when the primary direct link is probabilistically in outage or always in outage, respectively. For the secondary access, we assume that the secondary users utilize probabilistic time-division multiple-access (TDMA) scheme. We obtain the optimal assignment probabilities of the TDMA system under the QoS satisfaction of all the secondary users.

{\it Notation:} Throughout this paper, we use the following standard notation. The superscript $\dagger$ stands for the complex-conjugate transpose of a matrix or vector. $y^t$ denotes the
transposition of $y$. The symbols $||Y||$ and $|y|$ denote
the Euclidean norm of a vector $Y$ and the magnitude of a complex number
$y$, respectively. $\Pr \{.\}$ denotes the probability of the argument event. The notation ${\bf E}$ denotes the cardinality of the set $E$. Finally, $\overline{\mathcal{B}}=1-\mathcal{B}$. The main symbols of this paper are provided in Table \ref{table1}.

\begin{table}
\renewcommand{\arraystretch}{1}
\begin{center}
\begin{tabular}{ |c |l||  }
    \hline\hline
    Symbol & Notation \\[5pt]\hline
       ${\rm s}$ & {\footnotesize Secondary source} \\[5pt]\hline
          ${\rm p}$ & {\footnotesize Primary source} \\[5pt]\hline
           ${\rm sd}$ & {\footnotesize Secondary destination} \\[5pt]\hline
            ${\rm pd}$ & {\footnotesize Primary destination} \\[5pt]\hline
  $T$ & {\footnotesize Slot duration} \\[5pt]\hline
    $\mathcal{S}$ & {\footnotesize Set of relays} \\[5pt]\hline
        $\mathcal{M}$ & {\footnotesize Number of secondary users} \\[5pt]\hline
      $\Lambda$ & {\footnotesize Set of decoding relays} \\[5pt]\hline
    ${\bf \Lambda}$ & {\footnotesize Cardinality of decoding set or the number of decoding relays} \\[5pt]\hline
     $\mathcal{N}_\circ$ & {\footnotesize Variance of the AWGN at a receiving node in Watts/Hz} \\[5pt]\hline
       & {\footnotesize Transmit power of secondary users}\\  $P_{\rm s}$ & {\footnotesize for transmission of their own packets in Watts/Hz}\\[5pt]\hline
       & {\footnotesize Primary and secondary transmit power}\\ $P$  &  {\footnotesize while transmitting a primary packet in Watts/Hz} \\[5pt]\hline
\end{tabular}
\caption{List of main symbols.}
\label{table1}
\end{center}
\end{table}
\section{System Model}
In this paper, we assume a cognitive setting with one primary user and a set of secondary terminals with cardinality $\mathcal{M}$ secondary users. The set of secondary nodes is denoted by $\mathcal{S}=\{1,2,\dots,\mathcal{M}\}$. The secondary terminals are numbered $1,2,\dots,\mathcal{M}$. The secondary users share the spectrum using TDMA. Thus, each time slot one of the secondary users is scheduled for transmission. The probability of assigning user $v\in \mathcal{S}$ for transmission is $\omega_v\in[0,1]$. The secondary user scheduled for transmission is denoted by $v$. Time is slotted and a slot time is of length $T$ second. All secondary transmissions are assumed
to be slot synchronized \cite{liu2012cooperative}. All users are assumed to be always backlogged with data packets. In a given time slot, one of the secondary users is assigned for transmission, and the remainder of the secondary users operate as relay stations for the primary source. For sake of convenience, we refer to the secondary user scheduled for transmission as secondary source, the remainder of the secondary users as secondary relays, and finally, the destination of the secondary source and primary source as secondary and primary destinations, respectively.

We consider two cases based on the state of connectivity of the primary direct link. In the first case, we assume the existence of a direct link between the primary source and its destination. This link can be in outage with certain probability according to the transmission rate and link capacity. In the second case, we assume that the link between the primary source and its destination is always in outage, i.e., disconnected. The latter case happens when the distance between the primary source and its destination is large or the direct link is in deep shadowing due to surrounding physical obstacles.

In the first case, the proposed protocol is described as follows. The time slot is divided equally into two phases: $[0,T/2]$ and $[T/2,T]$. During $[0,T/2]$, the primary user broadcasts its packet to its destination and the secondary relays. The secondary relays attempt to decode the primary packet. We denote the set of secondary users that successfully decoded primary packet and will relay it as $\Lambda$, where $\Lambda \subseteq \mathcal{S}= \{1,2,\dots,\mathcal{M}\}$ and $v \notin \Lambda$. Thus, the cardinality of $\Lambda$ can take any integer value between $0$ and $\mathcal{M}-1$. Precisely, ${\bf \Lambda}=K \in\{0,1,2,\dots,\mathcal{M}-1\}$. During $[T/2,T]$, if $K\ge2$, the secondary relays forward the decoded primary packet to the primary destination. At the same time, the secondary user scheduled for transmission, user $v$, transmits its own packet. The secondary relays use a beamforming technique that nulls their interference at the destination of the user scheduled for transmission. If $K<2$, the secondary relays remain idle and the secondary source transmits its packet solely.
 At the end of the time slot, the primary receiver combines the received packets from the primary source and the secondary relays using Maximal Ratio Combining (MRC) technique.

 In the second case, since there is no direct link between the primary source and its destination, it is more appropriate to split the time slot into two unequal partitions. Specifically, we assume that the time slot is divided into $\zeta$ and $1-\zeta$ for the primary and secondary transmissions, respectively. We provide the details of both cases and prove the diversity-multiplexing tradeoff in each case.

In the proposed systems, the secondary relays utilize
the typical Decode-and-Forward (DF) relaying technique. In particular, the primary source broadcasts a packet to potential
relays and its destination. When more than one relay can decode the primary packet, the secondary relays that can successfully decode
the packet, then
forward the packet to the primary destination. The secondary relays
that could not decode this packet remain idle till the end of the time slot.

Wireless links exhibit fading and are corrupted by additive
white Gaussian noise (AWGN). We denote the
channel coefficient from node $\ell_1$ to node $\ell_2$ by $h_{\ell_1,\ell_2}\in \mathbb{C}$, where $\mathbb{C}$ denotes the set of all complex numbers. Here, $\ell_1\in\{{\rm p},1,2,3,\dots,\mathcal{M}\}$ and $\ell_2\in\{{\rm pd,sd},1,2,3,\dots,\mathcal{M}\}$, where $\ell_1\ne \ell_2$ and $\ell_2\ne v$, ${\rm p}$ denotes the primary source, and ${\rm pd}$ and ${\rm sd}$ denote the primary and secondary destinations, respectively. The fading is assumed to be
stationary with frequency non-selective Rayleigh block fading. The channel coefficient $h_{\ell_1,\ell_2}$ is assumed to be independent and identically distributed (i.i.d.) circularly symmetric complex Gaussian random variable with zero mean and unit
variance, i.e., $h_{\ell_1,\ell_2}\in \mathcal{CN}(0,1)$. That is, $h_{\ell_1,\ell_2}$ remains constant
during one time slot, and varies independently
from slot to slot. The thermal noise at any of the receiving nodes is assumed to be AWGN with zero mean and power spectral density $\mathcal{N}_\circ$ Watts/Hz. The primary and the secondary transmit power while transmitting a primary packet is $P$ Watts/Hz, whereas the secondary transmit power for its own data transmission is $P_{\rm s}$ Watts/Hz.

By assigning the beamforming weight ${g}^\dagger_k$, where ${g}^\dagger_k$ is conjugate of ${g}_k$, at each decoding relay $k\in \Lambda$, the received signal at the primary destination, ${\rm pd}$, from forwarding primary transmission by the relays when ${\bf \Lambda}=K\ge 2$ is given by
\begin{equation}
{r_{{\rm pd}}}\!=\!{g}^\dagger h_{\rm pd}^{\left(\Lambda\right)} \tilde x_{\rm p}\!+\!w_{\rm s}\!+\!z_{{\rm pd}}
\end{equation}
where $h_{\rm pd}^{\left(\Lambda\right)}\!=\![h_{1,{\rm pd}},h_{2,{\rm pd}},\dots,h_{K,{\rm pd}}]^t\in \mathbb{C}^K$ is  coefficient vector of channels from the decoding relays to the primary destination, $g\!=\![g_1,\dots,g_K]^t$ is the beamforming weight vector, $\tilde x_{\rm p}$ is the transmitted scalar signal with power $P$ Watts/Hz, $w_{\rm s}=h_{v,{\rm pd}} \tilde x_{\rm s}$ indicates the interference from the secondary source to the primary destination, $h_{v,{\rm pd}}$ is the channel coefficient between the secondary source and the primary destination, $\tilde x_{\rm s}$ is the transmitted secondary signal with power $P_{\rm s}$ Watts/Hz, and $z_{{\rm pd}}$ denotes the AWGN at the primary destination with variance $\mathcal{N}_{\circ}$.

Let $\alpha_{v,{\rm pd}}\!=\!|h_{v,{\rm pd}}|^2$. The instantaneous secondary interfering power at the primary
destination is $P_{\rm s}\alpha_{v,{\rm pd}}$.
 The instantaneous received {\rm SINR} at the primary destination from forwarding primary transmission by the relays is then given by
\begin{equation}
{\rm SINR}_{{\rm pd}}\!=\!\frac{|g^\dagger h_{\rm pd}^{\left(\Lambda\right)}|^2 P}{\mathcal{N}_{\circ}\!+\!P_{\rm s}\alpha_{v,{\rm pd}}}
\end{equation}

Note that the
interference from the secondary relays to the secondary source is eliminated due to the use of {\it zeroforcing beamforming
(ZFBF)}. On the contrary, the interference from secondary source to the
primary destination cannot be avoided. Next, we investigate the optimal ZFBF
weight vector.
In this paper, we use cooperative beamforming to obtain
cooperative diversity gain for the primary source while completely eliminating the
interference to the secondary source. Therefore, the optimal ZFBF
weight vector $g$ should be designed to maximize ${\rm SINR}_{{\rm pd}}$ and satisfy
$|{g}^\dagger h_{\rm sd}^{\left(\Lambda\right)}| \!=\! 0$, where $h_{\rm sd}^{\left(\Lambda\right)}=[h_{1,{\rm sd}},h_{2,{\rm sd}},\dots,h_{K,{\rm sd}}]^t \in \mathbb{C}^K$ denotes the coefficients from the decoding relays to the secondary destination, at the same time. Moreover, $g$ is
normalized to meet the power limit requirement at the relays.
In this context, the optimal weight vector, $g$, is exactly the
optimal solution of the following optimization problem:

\begin{eqnarray}
\label{opt2}
\begin{split}
    \underset{g}{\max.} & \ \ \ \ |g^\dagger h_{\rm pd}^{\left(\Lambda\right)}|^2,\ {\rm s.t.}   \ \ \ \   |{g^*}^\dagger h_{\rm sd}^{\left(\Lambda\right)}| \!=\! 0, \ ||g||\!=\!1
    \end{split}
\end{eqnarray}

Let $V$ be the subspace spanned by the channel coefficient vectors $h_{\rm sd}^{\left(\Lambda\right)}$. From (\ref{opt2}), the vector $g$ is orthogonal to $h_{\rm sd}^{\left(\Lambda\right)}$. Hence, it is perpendicular to each vector in $V$,
and thus belongs to $V^\perp$, where $V^\perp$ is the orthogonal complementary
subspace of $V$. In order to maximize $|g^\dagger h_{\rm pd}^{\left(\Lambda\right)} |^2$ in (\ref{opt2}),
we need to find the optimal vector $g \in V^\perp$ which is closest
to $h_{\rm pd}^{\left(\Lambda\right)}$. Using the results from Closest Point Theorem \cite{meyer2000matrix}, $g$ is the orthogonal
projection of $h_{\rm pd}^{\left(\Lambda\right)}$ onto the subspace $V^\perp$. Thus, $g^*= \Psi h_{\rm pd}^{\left(\Lambda\right)}$, where $\Psi$ denotes the orthogonal projector onto the subspace $V^\perp$. From the constraint $|| g ||= 1$, the optimal
solution can be given by $g^*= \frac{\Psi h_{\rm pd}^{\left(\Lambda\right)}}{||\Psi h_{\rm pd}^{\left(\Lambda\right)}||}$. The matrix $\Psi$ is given by $\Psi~=~I~-~h^{\left(\Lambda\right)}_{\rm sd}({h^{\left(\Lambda\right)}_{\rm sd}}^\dagger h^{\left(\Lambda\right)}_{\rm sd})^{\!-\!1} {h^{\left(\Lambda\right)}_{\rm sd}}^\dagger$, where $I$ denotes the identity matrix with size ${\bf \Lambda}\times {\bf \Lambda}$. The size of the projection matrix is ${\bf \Lambda} \times {\bf \Lambda}$.

In order to evaluate the performance when cooperative
beamforming is applied, it is necessary to obtain the distribution
of the channel gain $\alpha$. We will present it in the following
theorem \cite{liu2012cooperative}.

\begin{theorem}
If $h_{\rm pd}^{\left(\Lambda\right)}$ and $h_{\rm sd}^{\left(\Lambda\right)} \in \mathcal{ CN} (0, I)$ where $I$ denotes the identity matrix of size ${\bf \Lambda}\times {\bf \Lambda}$, the random
variable $\alpha \!=\!|{g^*}^\dagger h_{\rm pd}^{\left(\Lambda\right)}|^2$ is Chi-square distributed
with $2(K\!-\!1)$ degrees of freedom. Its probability density function (pdf) is characterized by
\begin{equation}
f_\alpha (x)\!=\!\frac{1}{(K\!-\!2)!} x^{K\!-\!2}\exp(\!-\!x), x\ge 0
\end{equation}
where $\mathcal{T}!$ is factorial of $\mathcal{T}$.
\end{theorem}
The proof of this theorem is found in \cite{liu2012cooperative}.

\section{Outage Probabilities and Diversity-Multiplexing Tradeoff}
Let $b$ denote the packets size and $W$ denotes the transmission bandwidth. Also, let the transmission time of node $\ell_1$ be $T_{\ell_1}$. The data rate for node $\ell_1$ is then given by $\mathcal{R}_{\ell_1}=b/W/T_{\ell_1}$ bits/sec/Hz. An outage of a link occurs if $\mathcal{R}_{\rm \ell_1}$ exceeds
the link capacity $C_{\ell_1,\ell_2}$.
\subsection{First Case: With Primary Direct Link}
When the primary source broadcasts a packet at a data rate $\mathcal{R}_{\rm p}$, a
relay $k$ becomes a decoding relay if the channel capacity
$C_{{\rm p},k} \ge \mathcal{R}_{\rm  p}$. The channel capacity $C_{{\rm p},k}$ is
given by $C_{{\rm p},k} \!=\! \log_2(1 \!+\! \gamma|h_{{\rm p},k}|^2)$, where
$\gamma\!=\! P/\mathcal{N}_\circ$ is the average transmitted signal-to-noise-ratio (SNR), $|h_{{\rm p},k}|^2$
is the channel power gain which is exponentially distributed
under Rayleigh fading. The probability of $k\in \Lambda$ is equal
to ${\rm Pr}\{k \in \Lambda\} \!=\! {\rm Pr}\{C_{{\rm p},k} \ge \mathcal{R}_{\rm  p}\} \!=\! \mathcal{L}\!=\!\exp(\!-\!  (2^{\mathcal{R}_{\rm  p}}\!-\!1)/\gamma)$.

For the primary source, the outage occurs in either one of the following events: 1) If the combined signal of the direct and the relaying links is undecodable at the primary destination; or 2) if the number of decoding relays is less than two relays, i.e., if ${\bf \Lambda}=K<2$, and the link between the primary source and its destination is in outage.

For the first outage event, the optimal {\rm SINR} at the primary destination is $\frac{|{g^*}^\dagger h_{\rm pd}^{\left(\Lambda\right)}|^2 P}{\mathcal{N}_{\circ}\!+\!P_{\rm s}\alpha_{v,{\rm pd}}}+\! \gamma|h_{{\rm p},{\rm pd}}|^2$. Let $\mathcal{R}=b/T/W$, hence, $\mathcal{R}_{\rm p}=2b/T/W=2\mathcal{R}$ bits/sec/Hz. The probability of outage due to the first event is given by $\Pr\{\frac{|{g^*}^\dagger h_{\rm pd}^{\left(\Lambda\right)}|^2 P}{\mathcal{N}_{\circ}\!+\!P_{\rm s}\alpha_{v,{\rm pd}}}\!+\! \gamma|h_{{\rm p},{\rm pd}}|^2 \le 2^{2\mathcal{R}}\!-\!1\}$.
Let $\eta=\frac{P}{\mathcal{N}_{\circ}\!+\!P_{\rm s}\alpha_{v,{\rm pd}}}$.
We have $\Pr\{{\bf \Lambda}=K\}\!=\!(\!\begin{array}{c}
  \mathcal{M}\!-\!1 \\
  K
\end{array}\!)\! \mathcal{L}^K  \overline{\mathcal{L}}^{(\mathcal{M}\!-\!K\!-\!1)}$. For a given interference channel gain $h_{v,{\rm pd}}$, primary direct link realization $h_{\rm p,pd}$, and the decoding relays set $\Lambda$ with cardinality ${\bf \Lambda} =K\ge 2$, the failure probability of the primary packet decoding is given by

\begin{equation}
\begin{split}
\Pr\{ |{g^*}^\dagger h_{\rm pd}^{\left(\Lambda\right)}|^2  +\frac{\gamma|h_{{\rm p},{\rm pd}}|^2}{\eta}<\frac{2^{2\mathcal{R}}\!-\!1}{\eta}|\Lambda,h_{\rm p,pd},\alpha_{v,{\rm pd}}\}
\end{split}
\end{equation}
This can be rewritten as
\begin{equation}
\begin{split}
\Pr\{ |{g^*}^\dagger h_{\rm pd}^{\left(\Lambda\right)}|^2  <\frac{2^{2\mathcal{R}}\!-\!1-\gamma|h_{{\rm p},{\rm pd}}|^2}{\eta}|\Lambda,h_{\rm p,pd},\alpha_{v,{\rm pd}}\}
\end{split}
\end{equation}
Using the fact that $ |{g^*}^\dagger h_{\rm pd}^{\left(\Lambda\right)}|^2$ is $2(K\!-\!1)$ Chi-square random variable, we get
\begin{equation}
\begin{split}
& \Pr\{ |{g^*}^\dagger h_{\rm pd}^{\left(\Lambda\right)}|^2  <\frac{2^{2\mathcal{R}}\!-\!1-\gamma|h_{{\rm p},{\rm pd}}|^2}{\eta}|\Lambda,h_{\rm p,pd},\alpha_{v,{\rm pd}}\}\\& \,\,\,\,\,\,\,\ \!=\!\int_0^\mathcal{X} f_\alpha (x) dx\!=\!1\!-\!\sum_{m\!=\!0}^{K\!-\!2}\frac{1}{m!}\mathcal{X}^m\exp(\!-\!\mathcal{X})
\end{split}
\end{equation}
where $\mathcal{X}\!=\!\frac{2^{2\mathcal{R}}\!-\!1-\gamma|h_{{\rm p},{\rm pd}}|^2}{\eta}\ge 0$, $\eta\!=\!\gamma/(1\!+\!\phi)$ and $\phi=\gamma_{\rm s}\alpha_{v,{\rm pd}}$. The positivity of $\mathcal{X}$ implies that $\mathcal{X}\!=\!\frac{2^{2\mathcal{R}}\!-\!1-\gamma|h_{{\rm p},{\rm pd}}|^2}{\eta}>0$; hence, $\frac{2^{2\mathcal{R}}\!-\!1}{\gamma}\ge |h_{{\rm p},{\rm pd}}|^2$. Note that if $\mathcal{X}$ is negative, there is no outage.

Averaging over the decoding set, the first outage probability for a fixed $h_{\rm p,pd}$ and $\alpha_{v,{\rm pd}}$ is then given by
\begin{equation}
\begin{split}
\label{roman}
&\Pr\{ |{g^*}^\dagger h_{\rm pd}^{\left(\Lambda\right)}|^2  <\mathcal{X}|h_{\rm p,pd},\alpha_{v,{\rm pd}}\}\\&\!=\! \sum_K\Pr\{ |{g^*}^\dagger h_{\rm pd}^{\left(\Lambda\right)}|^2  <\mathcal{X}|\Lambda,h_{\rm p,pd},\alpha_{v,{\rm pd}}\}\Pr\{{\bf \Lambda}=K\}\\& \,\,\,\,\,\,\,\ \!=\!\sum_{K\!=\!2}^{\mathcal{M}\!-\!1}\!(\!\begin{array}{c}
  \mathcal{M}\!-\!1 \\
  K
\end{array}\!)\! \mathcal{L}^K  \overline{\mathcal{L}}^{(\mathcal{M}\!-\!K\!-\!1)} (1\!-\!\sum_{m\!=\!0}^{K\!-\!2}\frac{1}{m!}\mathcal{X}^m\exp(\!-\!\mathcal{X}))
\end{split}
\end{equation}
Note that the above formula is valid due to the independency of the given events. Averaging over $\alpha_{\rm p,pd}=|h_{\rm p,pd}|^2$, we get

\begin{equation}
\label{poorr2}
\begin{split}
\nu_1&=\!\sum_{K\!=\!2}^{\mathcal{M}\!-\!1}\!(\!\begin{array}{c}
  \mathcal{M}\!-\!1 \\
  K
\end{array}\!)\! \mathcal{L}^K  \overline{\mathcal{L}}^{(\mathcal{M}\!-\!K\!-\!1)}\\&\,\,\,\,\,\,\,\,\,\  \times \Bigg[1-\sum_{m\!=\!0}^{K\!-\!2}\frac{1}{m!} \int_{0}^{\mathcal{Q}}\mathcal{X}^m\exp(\!-\!\mathcal{X}) \exp(-\alpha_{\rm p,pd})d \alpha_{\rm p,pd}\Bigg]
\end{split}
\end{equation}
 where $\mathcal{\mathcal{Q}}=\frac{2^{2\mathcal{R}}\!-\!1}{\gamma}$ and $\mathcal{X}\!=\!(\mathcal{\mathcal{Q}}-\alpha_{{\rm p},{\rm pd}}) (1+\phi)$. Let $\mathcal{W}=\int_{0}^{\mathcal{Q}}\mathcal{X}^m\exp(\!-\!\mathcal{X}) \exp(-\alpha_{\rm p,pd})d \alpha_{\rm p,pd}$. After some change of variables and algebra, we get the following:

\begin{equation}
\begin{split}
\mathcal{W}&\!=\!\exp(-\mathcal{Q})\frac{(1+\phi)^m}{\phi^{m+1}}  \!\int_{0}^{\mathcal{Q} \phi} \!R^m \exp(-R) d R
\\& \!=\exp(-\mathcal{Q})\frac{(1+\phi)^m}{\phi^{m+1}} \mathbb{L}(m+1,\mathcal{Q}\phi)
\label{tota}
\end{split}
\end{equation}
where $\mathbb{L}(m+1,s)=\int_{0}^s R^m \exp(-R) d R$ is the lower incomplete Gamma function. The outage probability $\nu_1$ for a given $\alpha_{v, {\rm pd}}$ (or $\phi$) is then given by

\begin{equation}
\label{poorr2}
\begin{split}
\nu_1&=\!\sum_{K\!=\!2}^{\mathcal{M}\!-\!1}\!(\!\begin{array}{c}
  \mathcal{M}\!-\!1 \\
  K
\end{array}\!)\! \mathcal{L}^K  \overline{\mathcal{L}}^{(\mathcal{M}\!-\!K\!-\!1)} \Bigg[1-\sum_{m\!=\!0}^{K\!-\!2}\frac{1}{m!}\\&\,\,\,\,\,\,\,\,\,\ \exp(-\mathcal{Q})\frac{(1+\phi)^m}{\phi^{m+1}} \mathbb{L}(m+1,\mathcal{Q}\phi)\Bigg]
\end{split}
\end{equation}

Consider the second outage event. The second outage event occurs when ${\bf \Lambda}\!=\!K<2$ and the link ${\rm p\rightarrow pd}$ is in outage. In this case, the outage probability is given by
\begin{equation}
\begin{split}
\label{pogy}
  \nu_2\!&=\!\Bigg[\sum_{K=0}^{1}\!(\!\begin{array}{c}
  \mathcal{M}\!-\!1 \\
  K
\end{array}\!)\! \mathcal{L}^K  \overline{\mathcal{L}}^{(\mathcal{M}\!-\!K\!-\!1)} \Bigg] \Pr\{|h_{\rm p,pd}|^2< \mathcal{Q}\}\\ & \!=\!\Bigg[\sum_{K=0}^{1}\!(\!\begin{array}{c}
  \mathcal{M}\!-\!1 \\
  K
\end{array}\!)\! \mathcal{L}^K  \overline{\mathcal{L}}^{(\mathcal{M}\!-\!K\!-\!1)} \Bigg] \overline{\mathcal{L}}\!=\!\sum_{K=0}^{1}\!(\!\begin{array}{c}
  \mathcal{M}\!-\!1 \\
  K
\end{array}\!)\! \mathcal{L}^K  \overline{\mathcal{L}}^{(\mathcal{M}\!-\!K)}
\end{split}
\end{equation}
The multiplication of the marginal probabilities to get the joint probability in (\ref{pogy}) is due to the independency of the channels gains. Summing up the outage probabilities, we obtain the following quantity for a given $\alpha_{v,{\rm pd}}$:
\begin{equation}
\begin{split}
& \nu_\phi\!=\!\nu_1\!+\! \nu_2\!=\!1\!-\!\sum_{K\!=\!2}^{\mathcal{M}\!-\!1}\!(\!\begin{array}{c}
  \mathcal{M}\!-\!1 \\
  K
\end{array}\!)\! \mathcal{L}^K  \overline{\mathcal{L}}^{(\mathcal{M}\!-\!K\!-\!1)} \\&\,\,\,\,\,\,\,\,\,\,\,\,\,\,\,\,\,\,\,\,\,\,\,\,\,\,\,\,\,\,\,\,\,\,\,\,\,\ \times \Bigg[\sum_{m\!=\!0}^{K\!-\!2}\frac{1}{m!} {\exp(-\mathcal{Q})}\frac{(1+\phi)^m}{\phi^{m+1}} \mathbb{L}(m+1,\mathcal{Q}\phi)\Bigg]
\end{split}
\end{equation}
Averaging over $\phi=\gamma_{\rm s} \alpha_{v,{\rm pd}}$, we get
\begin{equation}
\begin{split}
& \nu=\!1\!-\!\sum_{K\!=\!2}^{\mathcal{M}\!-\!1}\!(\!\begin{array}{c}
  \mathcal{M}\!-\!1 \\
  K
\end{array}\!)\! \mathcal{L}^K  \overline{\mathcal{L}}^{(\mathcal{M}\!-\!K\!-\!1)} \Bigg[\sum_{m\!=\!0}^{K\!-\!2}\frac{1}{m!}\\&\,\,\,\,\,\,\,\,\,\ \times \frac{\exp(-\mathcal{Q})}{\gamma_{\rm s}}\int_0^\infty\frac{(1+\phi)^m}{\phi^{m+1}} \mathbb{L}(m+1,\mathcal{Q}\phi) \exp(-\frac{\phi}{\gamma_{\rm s}})\ d\phi\Bigg]
\end{split}
\end{equation}
  In the sequel of this subsection, we approximate the primary outage probability, $\nu$, at high SNR, $\gamma$. At high $\gamma$, the term $(1\!-\!\sum_{m\!=\!0}^{K\!-\!2}\frac{1}{m!}\mathcal{X}^m\exp(\!-\!\mathcal{X}))$ in (\ref{roman}) is approximated to

\begin{equation}
\begin{split}
1\!-\!\sum_{m\!=\!0}^{K\!-\!2}\frac{1}{m!}\mathcal{X}^m\exp(\!-\!\mathcal{X})\!\approx\! \frac{1}{(K\!-\!1)!} \mathcal{X}^{K\!-\!1}= \frac{(2^{\mathcal{R}_{\rm p}}\!-\!1)^{K\!-\!1}}{\eta^{K\!-\!1}(K\!-\!1)!}
\label{pota}
\end{split}
\end{equation}
Note that $\exp(-|h_{\rm p,pd}|^2)\!\approx\! 1$ and $\exp(-\mathcal{X})\!\approx\! 1$ over $|h_{\rm p,pd}|^2 \in[0,\mathcal{Q}]$ at high SNR.

Integrating (\ref{pota}) with respect to $|h_{\rm p,pd}|^2$, and recalling that the feasible range of $|h_{\rm p,pd}|^2$ is $[0,\mathcal{Q}]$, we get the following expression in terms of $\mathcal{X}$:

\begin{equation}
\label{ou}
\begin{split} \frac{1}{(K\!-\!1)!} \frac{1}{(1+\phi)}\int_{0}^{\mathcal{Q} (1+\phi)}\mathcal{X}^{K\!-\!1} d\mathcal{X}=\frac{1}{K!} {(1+\phi)^{K\!-\!1}}\mathcal{Q}^{K}
\end{split}
\end{equation}

Substituting with (\ref{ou}) into (\ref{roman}), and using the fact that at high $\gamma$, $(1-\mathcal{L})\approx\mathcal{Q}$ and $\mathcal{L}\approx1$, we get

\begin{equation}
\begin{split}\nu_1\!\approx\!\sum_{K\!=\!2}^{\mathcal{M}\!-\!1}\!(\!\begin{array}{c}
  \mathcal{M}\!-\!1 \\
  K
\end{array}\!)\!  \mathcal{Q}^{(\mathcal{M}\!-\!K\!-\!1)}  \frac{1}{K!} {(1+\phi)^{K\!-\!1}}\mathcal{Q}^{K}
\end{split}
\end{equation}

Rearranging the result, we get
\begin{equation}
\begin{split}\nu_1\!\approx\!\Big[\sum_{K\!=\!2}^{\mathcal{M}\!-\!1}\!(\!\begin{array}{c}
  \mathcal{M}\!-\!1 \\
  K
\end{array}\!)\!    \frac{1}{K!} {(1+\phi)^{K\!-\!1}}\Big]\mathcal{Q}^{\mathcal{M}\!-\!1}
\end{split}
\end{equation}

The second outage probability, $\nu_2$, in (\ref{pogy}) is approximated by the lowest exponent of $\overline{\mathcal{L}}$, i.e., the term associated with $K=1$.\footnote{At high SNR $\gamma$, the probability of one secondary relays decodes the primary packet is significantly higher than the probability that none of the secondary relays decode the primary packet.} That is,

\begin{equation}
\begin{split}
&  \nu_2\!\approx\!(\!\begin{array}{c}
  \mathcal{M}\!-\!1 \\
  1
\end{array}\!)\! \mathcal{L}  \overline{\mathcal{L}}^{\mathcal{M}\!-\!1} \!\approx\!(\!\begin{array}{c}
  \mathcal{M}\!-\!1 \\
  1
\end{array}\!)\! \mathcal{Q}^{\mathcal{M}\!-\!1}
\end{split}
\end{equation}
Summing up the approximated probabilities, we get

\begin{equation}
\begin{split}
&  \nu_\phi\!=\!\nu_1\!+\!\nu_2\!\approx\! \Bigg[\!\sum_{K\!=\!2}^{\mathcal{M}\!-\!1}\!(\!\begin{array}{c}
  \mathcal{M}\!-\!1 \\
  K
\end{array}\!)\!    \frac{1}{K!} {(1+\phi)^{K\!-\!1}}+ \!(\!\begin{array}{c}
  \mathcal{M}\!-\!1 \\
  1
\end{array}\!)\!\Bigg]\! \mathcal{Q}^{\mathcal{M}\!-\!1}
\label{bing}
\end{split}
\end{equation}
  The expected value of $(1+\phi)^{K-2}$ is given by
 \begin{equation}
 \begin{split}
 & \frac{1}{\gamma_{\rm s}}\!\int_0^\infty (1+\phi)^{K-1} \exp(-\phi) d\phi\\ & \,\,\,\,\,\,\,\,\,\,\,\,\,\,\,\ \!=\! \frac{\exp(1)}{\gamma_{\rm s}} \int_1^\infty R^{K\!-\!1} \exp(\!-R) dR\!=\! \frac{\exp(1)}{\gamma_{\rm s}} \mathbb{U}(K,1)
 \end{split}
 \end{equation}
 where $\mathbb{U}(m+1,s)=\int_{s}^\infty R^m \exp(-R) d R$ is the incomplete upper Gamma function. The expected value of $\nu_\phi$ is then given by
 \begin{equation}
\begin{split}
&  \nu\!\approx\! \Bigg[\!\sum_{K\!=\!2}^{\mathcal{M}\!-\!1}\!(\!\begin{array}{c}
  \mathcal{M}\!-\!1 \\
  K
\end{array}\!)\!   { \frac{\exp(1)}{\gamma_{\rm s}}} \frac{\mathbb{U}(K,1)}{K!} \!+\! \!(\!\begin{array}{c}
  \mathcal{M}\!-\!1 \\
  1
\end{array}\!)\!\Bigg]\! \mathcal{Q}^{\mathcal{M}\!-\!1}
\label{bing2}
\end{split}
\end{equation}
From (\ref{bing2}), we can see that
the cooperative diversity order is equal to $\mathcal{M}-1$. From \cite{tse}, the transmission scheme
achieves the multiplexing gain $r$ if the data rate satisfies
$\lim_{\gamma \rightarrow \infty} \frac{\mathcal{R}({\gamma})}{\log {\gamma}}\!=\! r$, and the diversity $d$ if the outage probability can be approximated by $\lim_{\gamma \rightarrow \infty} \frac{\log\nu({\gamma})}{\log {\gamma}}\!=\!-d$ at high $\gamma$. The diversity-multiplexing tradeoff $d(r)$
measures the tradeoff between the the capacity of data transmission and reliability of data reception. For the first case, the multiplexing-diversity tradeoff is given by

\begin{equation}
\begin{split}
 d(r)=-\lim_{\gamma \rightarrow \infty} \frac{\log\nu({\gamma})}{\log {\gamma}}\!=\! (1\!-\!2r) (\mathcal{M}\!-\!1)
\end{split}
\end{equation}
with $0\le r\le 1/2$. The maximum diversity gain is $\mathcal{M}\!-\!1$, whereas the maximum multiplexing gain is $1/2$.

\subsection{Second Case: With No Primary Direct Link}
When there is no primary direct link, splitting the time slot into two partitions $\zeta$ and $1\!-\!\zeta$ would enhance the performance. Since each terminal transmits a packet of size $b$, the transmission rate of the primary user is $b/(\zeta T)$ bits/sec, whereas the rate of a secondary terminal in either transmission or retransmission of packets is $b/(1\!-\!\zeta)/T$ bits/sec. According to the previous description, an outage takes place
when one of the following two mutually exclusive events
occurs. One is that a packet is correctly received by less than
two relays. The other is that the packet is successfully decoded
by more than or equal to two relays but cannot be correctly
received by the primary destination.

Given that the transmission data rate for the primary user is $\mathcal{R}_\zeta\!=\!b/W/(T\zeta)=\frac{\mathcal{R}}{\zeta}$ bits/sec/Hz, when the primary source broadcasts a packet at a data rate $\mathcal{R}_\zeta$ bits/sec/Hz, a
relay $k\in \mathcal{S}$ becomes a decoding relay if the channel capacity
$C_{{\rm p},k} \ge \mathcal{R}_\zeta$. The probability of $k\in \Lambda$ is equal
to ${\rm Pr}\{k \in \Lambda\} \!=\! {\rm Pr}\{C_{{\rm p},k} \ge \mathcal{R}_\zeta\} \!=\! \mathcal{L}_\zeta\!=\!\exp(\!-\!  (2^{\mathcal{R}_\zeta}\!-\!1)/\gamma)$. We have  $\Pr\{{\bf \Lambda}=K\}\!=\!(\!\begin{array}{c}
  \mathcal{M}\!-\!1 \\
  K
\end{array}\!)\! \mathcal{L}_\zeta^K  \overline{\mathcal{L}_\zeta}^{(\mathcal{M}\!-\!K\!-\!1)}$, where $\Big(\!\begin{array}{c}
               y \\
               x
             \end{array}\!\Big)$ denotes $y$ choose $x$. For a given $\alpha_{v,{\rm pd}}$ and decoding set $\Lambda$ with cardinality ${\bf \Lambda}=K\ge2$, the failure probability of the primary packet decoding is given by

\begin{equation}
\begin{split}
&\Pr\{|{g^*}^\dagger h_{\rm pd}^{\left(\Lambda\right)}|^2<\mathcal{X}_{\zeta}|\Lambda,\alpha_{v,{\rm pd}}\}\\& \,\,\,\,\,\,\,\ \!=\!\int_0^{\mathcal{X}_{\zeta}} f_\alpha (x) dx\!=\!1\!-\!\sum_{m\!=\!0}^{K\!-\!2}\frac{1}{m!}\mathcal{X}_{\zeta}^m\exp(\!-\!\mathcal{X}_{\zeta})
\end{split}
\end{equation}
where $\mathcal{X}_{\zeta}\!=\!\frac{1}{\eta} (2^{\mathcal{R}_{\overline{\zeta}}}\!-\!1)$.
\begin{equation}
\begin{split}
\nu_1&=\Pr\{|{g^*}^\dagger h_{\rm pd}^{\left(\Lambda\right)}|^2<\mathcal{X}_{\zeta}|\alpha_{v,{\rm pd}}\}\\&\!=\! \sum_K\Pr\{|{g^*}^\dagger h_{\rm pd}^{\left(\Lambda\right)}|^2<\mathcal{X}_{\zeta}|\Lambda,\alpha_{v,{\rm pd}}\}\Pr\{{\bf \Lambda}=K\}\\& \,\,\,\,\,\,\,\ \!=\!\sum_{K\!=\!2}^{\mathcal{M}\!-\!1}\!(\!\begin{array}{c}
  \mathcal{M}\!-\!1 \\
  K
\end{array}\!)\! \mathcal{L}_\zeta^K  \overline{\mathcal{L}_\zeta}^{(\mathcal{M}\!-\!K\!-\!1)} (1\!-\!\sum_{m\!=\!0}^{K\!-\!2}\frac{1}{m!}\mathcal{X}_{\zeta}^m\exp(\!-\!\mathcal{X}_{\zeta}))
\label{pot1}
\end{split}
\end{equation}

The second outage probability, i.e., when ${\bf \Lambda}=K<2$, is given by
\begin{equation}
\begin{split}
\nu_2= \!\sum_{K=0}^{1}(\!\begin{array}{c}
  \mathcal{M}\!-\!1 \\
  K
\end{array}\!) \mathcal{L}_\zeta^K  \!\overline{\mathcal{L}_\zeta}^{(\mathcal{M}\!-\!K\!-\!1)}
\end{split}
\end{equation}
Summing up the two outage probabilities, and for a given interference realization $\alpha_{v,{\rm pd}}$ from the link S-PD, we get
\begin{equation}
\begin{split}
\nu_{\phi}\!=\!1\!-\!\sum_{K\!=\!2}^{\mathcal{M}\!-\!1}\!(\!\begin{array}{c}
  \mathcal{M}\!-\!1 \\
  K
\end{array}\!)\! \mathcal{L}_\zeta^K  \overline{\mathcal{L}_\zeta}^{(\mathcal{M}\!-\!K\!-\!1)} \sum_{m\!=\!0}^{K\!-\!2}\frac{1}{m!}\mathcal{X}_{\zeta}^m\exp(\!-\!\mathcal{X}_{\zeta})
\end{split}
\end{equation}
Averaging over $\alpha_{v,{\rm pd}}$, we get the following formula:
\begin{equation}
\begin{split}
\nu \!&=\!1\!-\!\sum_{K\!=\!2}^{\mathcal{M}\!-\!1}\!(\!\begin{array}{c}
  \mathcal{M}\!-\!1 \\
  K
\end{array}\!)\! \mathcal{L}_\zeta^K  \overline{\mathcal{L}_\zeta}^{(\mathcal{M}\!-\!K\!-\!1)}\\& \,\,\,\,\,\,\,\,\,\,\,\,\,\,\,\,\,\,\,\ \times \Bigg[ \sum_{m\!=\!0}^{K\!-\!2}\frac{1}{m!}\frac{\exp(1/\gamma_{\rm s})}{\mathcal{Q}_\zeta \gamma_{\rm s}(1+\frac{1}{\mathcal{Q}_\zeta\gamma_{\rm s}})^{m+1}} \mathbb{U}(m+1,\mathcal{Q}_\zeta+\frac{1}{\gamma_{\rm s}})\Bigg]
\end{split}
\end{equation}
where $\mathcal{Q}_\zeta=\frac{2^{\mathcal{R}_{\overline{\zeta}}}-1}{\gamma}$.

When the average SNR, $\gamma$, is sufficiently high, the term $(1\!-\!\sum_{m\!=\!0}^{K\!-\!2}\frac{1}{m!}\mathcal{X}_{\zeta}^m\exp(\!-\!\mathcal{X}_{\zeta}))$ in (\ref{pot1}) is approximated to

\begin{equation}
\begin{split}
1\!-\!\sum_{m\!=\!0}^{K\!-\!2}\frac{1}{m!}\mathcal{X}_{\zeta}^m\exp(\!-\!\mathcal{X}_{\zeta})\!\approx\! \frac{1}{(K\!-\!1)!} \mathcal{X}_{\zeta}^{K\!-\!1}= \frac{(2^{\mathcal{R}_{\overline{\zeta}}}\!-\!1)^{K\!-\!1}}{\eta^{K\!-\!1}(K\!-\!1)!}
\end{split}
\end{equation}

We also have $ \mathcal{L}_\zeta\!\approx\! 1 $, $ \frac{1}{\gamma}\approx 0$, $\frac{2^{\mathcal{R}_{\overline{\zeta}}}-1}{\gamma}\!\approx\! \frac{2^{\mathcal{R}_{\overline{\zeta}}}}{\gamma}$, and $\frac{2^{\mathcal{R}_{\zeta}}-1}{\gamma}\!\approx\! \frac{2^{\mathcal{R}_{\zeta}}}{\gamma}$. Thus, the first outage probability is approximated as

\begin{equation}
\begin{split}
&\nu_1\approx\!\sum_{K\!=\!2}^{\mathcal{M}\!-\!1}\!(\!\begin{array}{c}
  \mathcal{M}\!-\!1 \\
  K
\end{array}\!)\!  (\frac{2^{\mathcal{R}_{{\zeta}}}\!-\!1}{\gamma})^{(\mathcal{M}\!-\!K\!-\!1)} \frac{(\frac{2^{\mathcal{R}_{\overline{\zeta}}}\!-\!1}{\gamma})^{K\!-\!1}(1+\phi)^{K\!-\!1}}{(K\!-\!1)!} \\& =\frac{1}{\gamma^{\mathcal{M}\!-\!2}}\sum_{K\!=\!2}^{\mathcal{M}\!-\!1}\!(\!\begin{array}{c}
  \mathcal{M}\!-\!1 \\
  K
\end{array}\!)\!  ({2^{\mathcal{R}_{{\zeta}}}})^{(\mathcal{M}\!-\!K\!-\!1)} ({2^{\mathcal{R}_{\overline{\zeta}}}})^{K\!-\!1}\frac{(1+\phi)^{K\!-\!1}}{(K\!-\!1)!}\\& =\frac{1}{\gamma^{\mathcal{M}\!-\!2}}\sum_{K\!=\!2}^{\mathcal{M}\!-\!1}\!(\!\begin{array}{c}
  \mathcal{M}\!-\!1 \\
  K
\end{array}\!)\!  ({2^{\mathcal{R}}})^{\frac{\mathcal{M}\!-\!K\!-\!1}{\zeta}} ({2^{\mathcal{R}}})^{\frac{K\!-\!1}{\overline{\zeta}}}\frac{(1+\phi)^{K\!-\!1}}{(K\!-\!1)!}
\end{split}
\end{equation}
After some modifications, we get

\begin{equation}
\begin{split}
\label{xor}
&\nu_1\!\approx\!\frac{1}{\gamma^{\mathcal{M}\!-\!2}}\sum_{K\!=\!2}^{\mathcal{M}\!-\!1}\!(\!\begin{array}{c}
  \mathcal{M}\!-\!1 \\
  K
\end{array}\!)\!  ({2^{\mathcal{R}}})^{\frac{1}{\zeta} (\!\mathcal{M}\!-\!1\!)-\frac{1}{\overline{\zeta}}} ({2^{\mathcal{R}}})^{({\frac{1}{\overline{\zeta}}}-\frac{1}{\zeta}) K}\frac{(1+\phi)^{K\!-\!1}}{(K\!-\!1)!}
\end{split}
\end{equation}
If ${\frac{1}{\overline{\zeta}}}-\frac{1}{\zeta}\ge 0$, i.e., $\frac{1}{\overline{\zeta}}\ge \frac{1}{\zeta}$ or ${\zeta} \ge 1/2$, then the dominant term of the summation corresponding to $K\!=\!K_{\max}\!=\!\mathcal{M}\!-\!1$. That is,
\begin{equation}
\begin{split}
\nu_1\!\approx\!(\!\begin{array}{c}
  \mathcal{M}\!-\!1 \\
  \mathcal{M}\!-\!1
\end{array}\!)\!  (\frac{2^{\frac{1}{\overline{\zeta}} \mathcal{R}}}{\gamma})^{(\mathcal{M}\!-\!2)} \frac{(1+\phi)^{\mathcal{M}\!-\!2}}{(\mathcal{M}\!-\!2)!}
\end{split}
\end{equation}

The second outage probability is approximated by the value in the summation with exponent $K=1$; hence, we have

\begin{equation}
\label{romansan}
\begin{split}
\nu_2\!&\!\approx\!  (\!\begin{array}{c}
  \mathcal{M}\!-\!1 \\
  1
\end{array}\!)   \overline{\mathcal{L}_\zeta}^{(\mathcal{M}\!-\!2)}\!\\& \,\,\,\,\,\,\,\,\,\ \approx \!(\!\begin{array}{c}
  \mathcal{M}\!-\!1 \\
  1
\end{array}\!)   (\!\frac{2^{\mathcal{R}_{\zeta}}\!-\!1}{\gamma})^{(\mathcal{M}\!-\!2)}\!\approx\!(\!\begin{array}{c}
  \mathcal{M}\!-\!1 \\
  1
\end{array}\!)   (\!\frac{2^{{\frac{1}{\zeta}}\mathcal{R}}}{\gamma})^{(\mathcal{M}\!-\!2)}
\end{split}
\end{equation}

Summing up the approximated probabilities, we get

\begin{equation}
\begin{split}
\nu_\phi\!=\!\nu_1\!+\!\nu_2\!\approx\!\Bigg[\!(\!\begin{array}{c}
  \mathcal{M}\!-\!1 \\
    \mathcal{M}\!-\!1
\end{array}\!)\!   \frac{(1+\phi)^{\mathcal{M}\!-\!2}}{(\mathcal{M}\!-\!2)!}+(\!\begin{array}{c}
  \mathcal{M}\!-\!1 \\
  1
\end{array}\!)  \! \Bigg](\!\frac{2^{{\frac{1}{\overline{\zeta}}}\mathcal{R}}}{\gamma})^{(\mathcal{M}\!-\!2)}
\label{pokra11}
\end{split}
\end{equation}
with $\zeta\ge 1/2$.
  The expected value of $(1+\phi)^{\mathcal{M}-2}$ is given by
 \begin{equation}
 \begin{split}
 & \frac{1}{\gamma_{\rm s}}\!\int_0^\infty (1+\phi)^{\mathcal{M}-2} \exp(-\phi) d\phi\\ & \,\,\,\,\,\,\,\,\,\,\,\,\,\,\,\ \!=\! \frac{\exp(1)}{\gamma_{\rm s}} \int_1^\infty R^{\mathcal{M}\!-\!2} \exp(\!-R) dR\!=\! \frac{\exp(1)}{\gamma_{\rm s}} \mathbb{U}(\mathcal{M}\!-\!1,1)
 \end{split}
 \end{equation}
 Therefore, the expected value of $\nu_\phi$ is given by
 \begin{equation}
\begin{split}
\nu_\phi\!\approx\!\Bigg[\! \frac{\frac{\exp(1)}{\gamma_{\rm s}} \mathbb{U}(\mathcal{M}-1,1)}{(\mathcal{M}\!-\!2)!}+(\!\begin{array}{c}
  \mathcal{M}\!-\!1 \\
  1
\end{array}\!)  \! \Bigg](\!\frac{2^{{\frac{1}{\overline{\zeta}}}\mathcal{R}}}{\gamma})^{(\mathcal{M}\!-\!2)}
\label{pokra1}
\end{split}
\end{equation}
 In a similar fashion, we can get the expressions for ${\zeta}\le1/2$. If ${\zeta} \le 1/2$, we substitute with $K=K_{\min}=2$ into (\ref{xor}). The approximated value of $\nu_1$ is then given by
\begin{equation}
\begin{split}
&\nu_1\!\approx\!\frac{1}{\gamma^{\mathcal{M}\!-\!2}}\!(\!\begin{array}{c}
  \mathcal{M}\!-\!1 \\
  2
\end{array}\!)\!  ({2^{\mathcal{R}}})^{\frac{1}{\zeta} (\!\mathcal{M}\!-\!1\!)-\frac{1}{\overline{\zeta}}} ({2^{\mathcal{R}}})^{2({\frac{1}{\overline{\zeta}}}-\frac{1}{\zeta}) }{(1+\phi)}
\end{split}
\end{equation}
Rearranging the equation, we get
\begin{equation}
\label{bokrax}
\begin{split}
&\nu_1\!\approx\!\frac{1}{\gamma^{\mathcal{M}\!-\!2}}\!(\!\begin{array}{c}
  \mathcal{M}\!-\!1 \\
  2
\end{array}\!)\!  ({2^{\mathcal{R}}})^{\frac{1}{\zeta} (\mathcal{M}\!-\!2)+\frac{1}{\overline{\zeta}}-\frac{1}{\zeta}} {(1+\phi)}
\end{split}
\end{equation}
Note that the second outage probability $\nu_2$ follows (\ref{romansan}). Recalling that $\frac{1}{\zeta}\ge \frac{1}{\overline{\zeta}}$ (or ${\zeta}<1/2$), we can see that the exponent of $\nu_2$ is greater than that of $\nu_1$. Hence, $\nu_2$ dominates $\nu_1$. The summation of the approximated outage probabilities is then given by
\begin{equation}
\begin{split}
\nu_\phi=\nu_1+\nu_2\!\approx\!(\!\begin{array}{c}
  \mathcal{M}\!-\!1 \\
  1
\end{array}\!)   (\!\frac{2^{{\frac{1}{\zeta}}\mathcal{R}}}{\gamma})^{(\mathcal{M}\!-\!2)}
\label{pokrax}
\end{split}
\end{equation}
with ${\zeta}<1/2$. Since the approximated value of $\nu_\phi$ is independent of $\phi$, $\nu=\nu_\phi$. 

From (\ref{pokrax}), we can see that
the cooperative diversity order is equal to $\mathcal{M}\!-\!2$.
The transmission scheme
achieves the multiplexing gain $r$ if the data rate satisfies
$\lim_{\gamma \rightarrow \infty} \frac{\mathcal{R}({\gamma})}{\log {\gamma}}\!=\! r $, and the diversity $d$ if the outage probability can be approximated by $\lim_{\gamma \rightarrow \infty} \frac{\log\nu({\gamma})}{\log {\gamma}}\!=\!-d$ at high $\gamma$.

From the preceding derivation, the diversity-multiplexing
tradeoff $d (r)$ can be computed as follows: if $\zeta\le 1/2$, substituting by $\mathcal{R}=r\log \gamma$ into (\ref{pokra1}) and taking the limit of $\gamma$ to infinity, we get

\begin{equation}
\begin{split}
d(r)\!&=\!-\!\lim_{\gamma\rightarrow \infty} \frac{\log \nu(\gamma)}{\log {\gamma}}\! =
 \!(1\!-\! \frac{r}{\zeta}) (\mathcal{M}\!-\!2)
\end{split}
\end{equation}
If $\zeta\ge 1/2$, substituting by $\mathcal{R}=r \log \gamma$ into (\ref{pokrax}) and taking the limit of $\gamma$, we get
\begin{equation}
\begin{split}
d(r)\!&=\!-\!\lim_{\gamma\rightarrow \infty} \frac{\log \nu(\gamma)}{\log {\gamma}}\! =
 \!(1\!-\!\frac{r}{\overline{\zeta}}) (\mathcal{M}\!-\!2)
\end{split}
\end{equation}

Combining both cases, we get
\begin{equation}
\begin{split}
d(r)\!&=
 \!(1\!-\!\min\{\frac{1}{\zeta},\frac{1}{1-\zeta}\} r) (\mathcal{M}\!-\!2)
\end{split}
\end{equation}
with $0\le r \le \min\{\zeta,1-\zeta\}$. The maximum achievable multiplexing gain is $\min\{\zeta,1-\zeta\}$ and the maximum achievable diversity gain is $\mathcal{M}\!-\!2$.
\section{Primary and Secondary Throughput}
According to the description of the proposed protocol, the secondary throughput in the $n$th case, $n\in\{1,2\}$ for the first and second cases, respectively, is given by
\begin{equation}
\begin{split}
\mu^{(n)}_{j}&\!=\!\omega_j f^{(n)}_{j}
\end{split}
\end{equation}
where $j\in\mathcal{S}$, $\omega_j$ is the probability of scheduling the $j$th secondary user for transmission and  $f^{(n)}_{\rm s}$ denotes the probability that the link connecting the secondary source scheduled for transmission and its destination being not in outage when the terminals operate under the $n$th case. For $n=1$, this probability is given by $f^{(1)}_{\rm s}=\Pr\{\mathcal{R}_{\rm s} \!\ge\! \log_2(1 \!+\! \gamma_{\rm s}|h_{\rm s,sd}|^2)\}$. Since $\mathcal{R}_{\rm s}=2\mathcal{R}$ when $n=1$,
\begin{equation}
\begin{split}
f^{(1)}_{\rm s}=\exp(-\frac{2^{2\mathcal{R}}-1}{\gamma_{\rm s}})
\end{split}
\end{equation}
where $\gamma_{\rm s}=P_{\rm s}/\mathcal{N}_\circ$.

In the second case, i.e., when the primary direct link is always in outage, $\mathcal{R}_{\rm s}= \frac{\mathcal{R}}{\overline{\zeta}}$; hence,
\begin{equation}
\begin{split}
f^{(2)}_{v}=\exp(-\frac{2^{\frac{\mathcal{R}}{\overline{\zeta}}}-1}{\gamma_{\rm s}})
\end{split}
\end{equation}

The primary throughput is given by $\mu^{(n)}_{\rm p}=1-\nu^{(n)}$, where the superscript `$n$' is added to distinguish between the studied cases. In this paper, we consider that each user has certain QoS requirement specified by a constraint on its throughput. Specifically, the PU throughput constraint is $\mu^{(n)}_{\rm p}=1-\nu^{(n)}\ge \lambda_{\rm p}$, whereas the $j$th secondary user throughput constraint is $\mu^{(n)}_{j}\ge \lambda_j$, where $\lambda_{\rm p}$ and $\lambda_j$ are the minimum required throughput for the PU and the $j$th secondary users, respectively. Under the first case, the optimal time resource assignments $\omega$'s can be obtained via finding the feasible set of the following linear constraints:
\begin{eqnarray}
\begin{split}
  \lambda_{\rm p}&\le \mu^{(1)}_{\rm p}=1-\nu^{(1)}, \\ \lambda_{j} &\le \mu^{(1)}_{j}=\omega_{j} \exp(-\frac{2^{{2\mathcal{R}}}-1}{\gamma_{\rm s}}) \forall j\in \mathcal{S}, \ \sum_{j\in \mathcal{S}} \omega_j = 1
    \end{split}
\end{eqnarray}
From the second constraint, $\lambda_{j} \le \mu^{(n)}_{j}$, we have

\begin{equation}
\begin{split}
\omega_j\ge \frac{\lambda_j}{\exp(-\frac{2^{{2\mathcal{R}}}-1}{\gamma_{\rm s}})} \forall j
\end{split}
\end{equation}
Summing both sides over $j \in \mathcal{S}$, and using the third constraint, we get
\begin{eqnarray}
\begin{split}
\sum_{j\in \mathcal{S}}\omega_j=1\ge \sum_{j\in \mathcal{S}}\frac{\lambda_j}{\exp(-\frac{2^{{2\mathcal{R}}}-1}{\gamma_{\rm s}})}
    \end{split}
\end{eqnarray}
Rearranging the result, we get
\begin{eqnarray}
\begin{split}
\sum_{j\in \mathcal{S}} {\lambda_j} \le{\exp(-\frac{2^{{2\mathcal{R}}}-1}{\gamma_{\rm s}})}
    \end{split}
\end{eqnarray}
with $  \lambda_{\rm p}\le 1-\nu^{(1)} $. The maximum for the $k$th secondary user for a given set of requirements $R_q=(\lambda_1,\lambda_2,\dots,\lambda_\mathcal{M})$, $\lambda_k\notin R_q$, for the other users is given by
\begin{eqnarray}
\begin{split}
{\lambda_k}={\exp(-\frac{2^{{2\mathcal{R}}}-1}{\gamma_{\rm s}})}-\sum_{\substack{{j\in \mathcal{S}}\\{j\ne k}}} {\lambda_j}
\label{pgo1}
    \end{split}
\end{eqnarray}

In this case, the optimal probabilities for time resource sharing among the SUs for a given $\mathcal{R}_n$ are given by
\begin{equation}
\begin{split}
\omega_j^*= \frac{\lambda_j}{\exp(-\frac{2^{{2\mathcal{R}}}-1}{\gamma_{\rm s}})} \forall j\ne k, \omega^*_k=1-\sum_{\substack{{j\in \mathcal{S}}\\{j\ne k}}} \omega_j^*
\end{split}
\end{equation}
with $  \lambda_{\rm p}\le 1-\nu^{(1)} $.

In the second proposed case, we add $\zeta$ to the optimization variables of the system. For a fixed $\zeta$, the constraints are linear and the optimal set is given by
   \begin{figure}
  \centering
  \includegraphics[width=1\columnwidth]{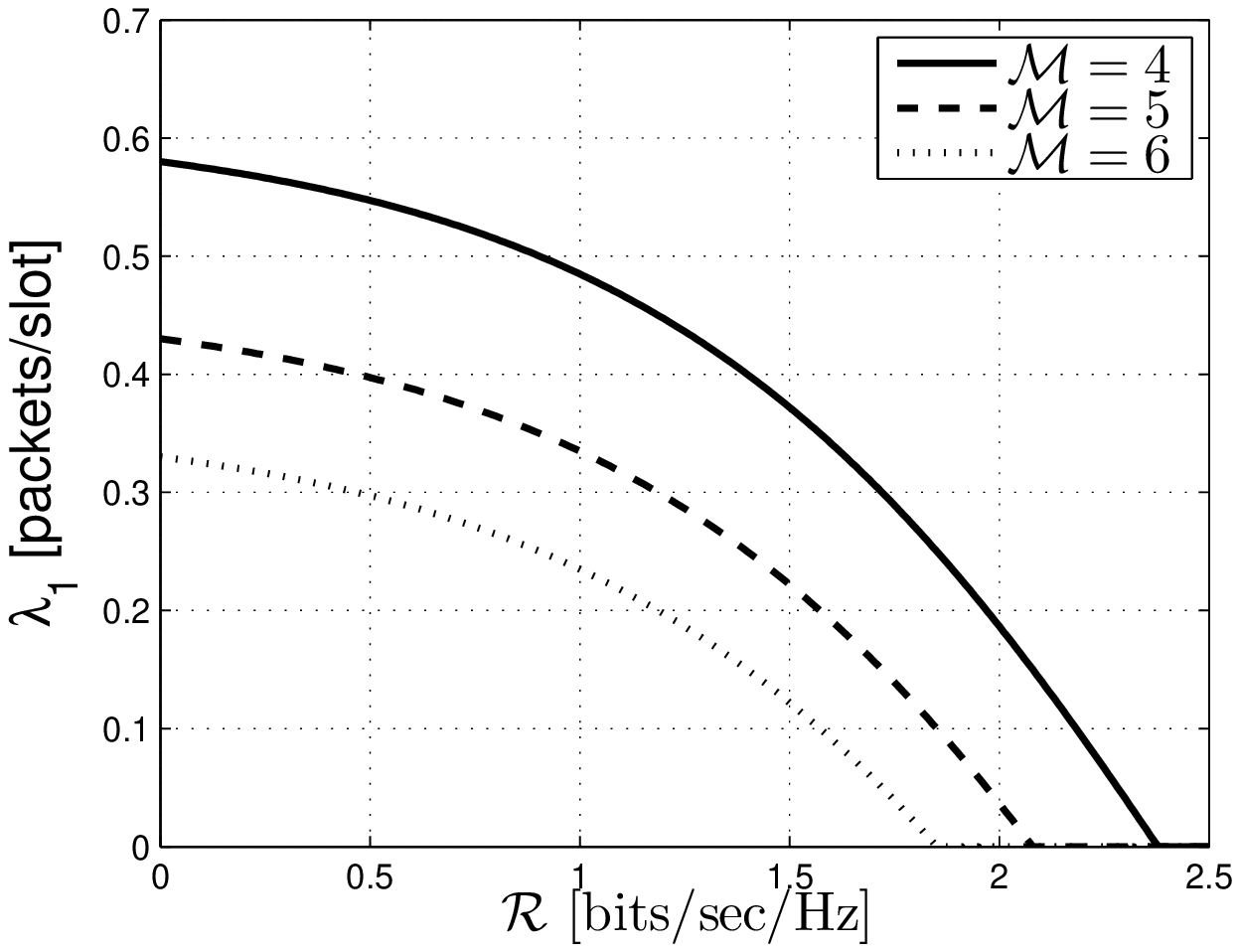}\\
   \caption{Maximum allowable QoS for secondary user $1$ when there is a primary direct link.}\label{fig1}
\end{figure}
\begin{eqnarray}
\begin{split}
  \sum_{j\in \mathcal{S}} {\lambda_j} \le{\exp(-\frac{2^{\frac{\mathcal{R}}{\overline{\zeta}}}-1}{\gamma_{\rm s}})}
    \end{split}
\end{eqnarray}
with $1-\nu^{(2)}\ge \lambda_{\rm p}$. The maximum throughput for the $k$th secondary user for a given set of requirements $R_q=(\lambda_1,\lambda_2,\dots,\lambda_\mathcal{M})$, $\lambda_k\notin R_q$, for the other users is given by
\begin{eqnarray}
\begin{split}
{\lambda_k}={\exp(-\frac{2^{\frac{\mathcal{R}}{\overline{\zeta}}}-1}{\gamma_{\rm s}})}-\sum_{\substack{{j\in \mathcal{S}}\\{j\ne k}}} {\lambda_j}
\label{pgo2}
    \end{split}
\end{eqnarray}
The optimal time resources among secondary users are

\begin{equation}
\begin{split}
\omega_j^*= \frac{\lambda_j}{\exp(-\frac{2^{\frac{\mathcal{R}}{\overline{\zeta}}}-1}{\gamma_{\rm s}})} \forall j\ne k, \omega^*_k=1-\sum_{\substack{{j\in \mathcal{S}}\\{j\ne k}}} \omega_j^*
\label{zetaeqn}
\end{split}
\end{equation}
with $1-\nu^{(2)}\ge \lambda_{\rm p}$. We note that the optimal values are parameterized by $\zeta$.
The optimal value of $\zeta$ is any value that satisfies the constraints.
\section{Numerical Results and Conclusions}
\begin{figure}
  \centering
  \includegraphics[width=1\columnwidth]{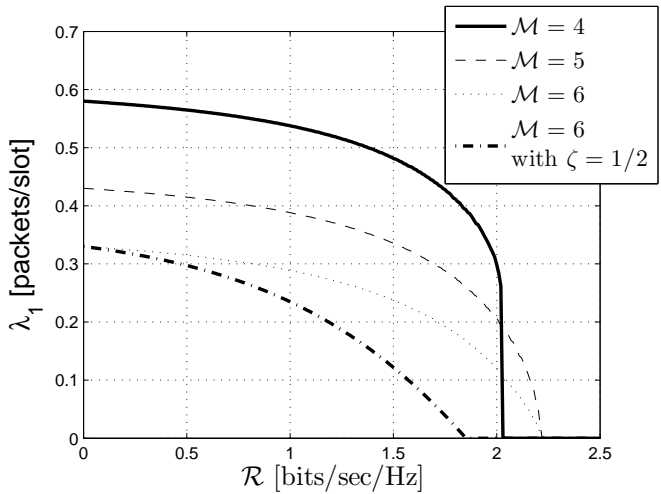}\\
   \caption{Maximum allowable QoS for secondary user $1$ when there is no primary direct link.}\label{fig2}
      \ 
\end{figure}

In this section, we provide some simulations for the proposed protocol. In Figs. \ref{fig1} and \ref{fig2}, we show the maximum allowable (supportable) QoS requirement for user $1$ for a given set of requirements for the other secondary users with and without primary direct link, respectively. The set of used parameters is: $\gamma=50$, $\gamma_{\rm s}=30$, $\lambda_2=0.1$ packets/slot,
$\lambda_3=0.2$ packets/slot, $\lambda_4=0.1$ packets/slot, $\lambda_5=0.15$ packets/slot, $\lambda_6=0.1$ packets/slot and $\lambda_{\rm p}=0.1$ packets/slot. As shown in the figure, increasing the number of secondary users, $\mathcal{M}$, decreases $\lambda_1$. This is because increasing $\mathcal{M}$ decreases the rate that one of the secondary users can get. This fact respects the constraints on the sum of requirements in (\ref{pgo1}) and (\ref{pgo2}).

As shown in  Fig. \ref{fig2}, increasing the number of secondary users increases the feasible range of $\mathcal{R}$. This is because increasing the secondary users increases the possibility of correct primary packet decoding by the secondary relays and, hence, increases the possibility of primary user satisfaction. Note that without cooperation the primary throughput when there is no primary direct link is zero. From the figures, we see the significant gain for the primary and secondary users under cooperation. In Fig. \ref{fig2}, we also plot the case of $\mathcal{M}=6$ with $\zeta=1/2$ to show importance of splitting the time slot unequally when the direct link of the primary user is always in outage. As shown in the figure, splitting the time slot can significantly improve the performance.

 In this paper, we have proposed a cooperative relaying protocol which involves cooperation among primary and secondary users. The secondary users aid each others to achieve certain QoS requirements simultaneously with the required QoS for the primary user. We have derived the optimal time slots assignments for secondary users. We have derived the diversity-multiplexing gain curves for the proposed systems.


\bibliographystyle{IEEEtran}
\balance
\bibliography{IEEEabrv,bibgc2014}
\end{document}